# EFFECT OF ENERGY LEAKAGE ON THE ENERGY RESOLUTION OF E.M. SAMPLING CALORIMETERS

*O.P. Gavrishchuk[1], V.E. Kovtun[2], T.V. Malykhina[2]*
[1]*Joint Institute for Nuclear Research (JINR), Dubna, Russia;*
[2]*V.N. Karazin Kharkiv National University, Kharkiv, Ukraine*
*E-mail: kovtun@univer.kharkov.ua*

The electromagnetic sampling calorimeters of the SPD experiment (NICA collider) are being investigated by Monte Carlo method. The simulation is used to study in detail the influence of energy leakages from the module on its energy resolution. The values of the stochastic and constant terms are obtained for the SPD calorimeter prototypes.

## INTRODUCTION

The Spin Physics Detector (SPD [1], NICA, JINR) experiment is designed to study spin-dependent parton distributions and correlations in nuclei at the NICA collider in beams of intense polarized relativistic ions.

One of the main detecting systems of the facility is an electromagnetic (EM) calorimeter (SPD ECal). The design is based on the KOPIO [2] prototype sampling calorimeter. Lead was chosen as an absorber, and a plastic scintillator was chosen as an active medium. The expected energy resolution in the energy range from 50 MeV to 16 GeV should be as good as possible, but not worse $\sim 5\%/\sqrt{E(\text{GeV})}$. However, subsequent refinements and discussion of the installation design [3] showed that the length of the entire module structure should not exceed 50 cm. Both requirements are difficult to reconcile. Therefore, a careful analysis of the factors affecting resolution is required.

The main optimization of the design of the calorimeter module is carried out in terms of energy resolution parameters. Usually the function of energy resolution versus energy is parameterized by the quadratic summation of two terms – stochastic term and constant term:

$$\frac{\sigma_E}{E} = \frac{a}{\sqrt{E}} \oplus b. \qquad (1)$$

The coefficient *a* have a statistical nature and represents a combination of Poisson-type effects – sampling, photostatistics fluctuations and electronic noise. The coefficient *b* dominates at high energies and is associated with the design features of the calorimeter module, inhomogeneity of sensitive elements, module dimensions, energy leakage, different calibrations of modules, inhomogeneities of light collection and light attenuation length in fibers, inaccuracies in the manufacture of absorbers, passive materials, etc.

In this work, we will consider the so-called ideal calorimeter, when all effects, except for sampling and energy leakage from the module, are artificially turned off. The task is to study the effect of energy leakage from the SPD calorimeter module which has a limited size. It has long been noticed and discussed many times that parameterization (1) is "fundamentally incorrect" [4, p. 48]. The energy resolution is fitted with function of several types. The ultimate goal is to obtain the parameters of several possible prototype calorimeters.

As will be seen from the results of our detailed simulations, the specific shape of the energy resolution terms will depend from shower energy leakage from the calorimeter, which is almost always present even in small amounts. The EM leakages may not be strongly affected on the shape and parameters of the total absorption peak. But the energy resolution is determined by the entire set of energy points and the deviation from the form of (1) already can be significant.

The approach is based on careful selection the energy spectra according to the Chi-square test. Next, we use a new energy resolution approximation function that explicitly takes into account energy leakages from a module of real dimensions and has good convergence to $\chi^2/ndf \sim 1$.

Specifically, the task was solved by studying the total energy resolution of the prototypes tower ECal SPD calorimeter depending on the number of layers and absorber thickness.

## 1. MONTE CARLO SIMULATION
### 1.1. IDEAL CASE

The influence of the effects of longitudinal and transverse EM energy leakages on the resolution of the electromagnetic sampling of the tile calorimeter is considered. Only the sampling of the shower EM fluctuations and the leakage of the shower energy from the module are taken into account. It's so-called ideal case. The effect hadronic interactions (nuclear excitations etc.) and nuclear produce particles (neutron, neutrinos etc.) is small.

For SPD purposes, the module is selected, which is a lead-scintillator sandwich structure with *L* pairs of plates (100, 120, 140, 160, 180, 200, 220, 240, 260, 280, 300, 500, 1000). The module *L*=2000 (denote as $L=\infty$) represents a "benchmark". In this case, longitudinal and lateral leakages are guaranteed to be absent, and the beam spreads from the center of the module. Transverse dimension of the calorimeter module SPD is equal $H_{yz}$=110 mm [1]. The absorber thickness was chosen to be $D_{Pb}$=0.3, 0.4, 0.5 mm. Scintillator thickness $D_S$=1.5 mm did not change because due to the manufacturing technology tiles for SPD ECal.

The simulation of the calorimeter module was performed using the Geant4.10.6 Monte Carlo simulation package [5]. The statistical error was at the level 1%.

Physics Lists: emstandard_opt3, $E_{cut}$ = 100 μm. The assembly of modules has the nonet structure (3×3) to better approximate the actual plant structure. The gun particle is directed to the central tower along X axis. Geometric parameters were varied to better identify the effects of leakages or to test a number of assumptions about energy leakages.

The total deposited EM energy was determined from the absorbed energy summation in the active medium from the $L$ plates of the module scintillator. A large number ($E_0(i)[i=1-60]$ points) was chosen of the input beam energy in the energy range 50 MeV…16 GeV to ensure high accuracy of energy resolution approximation. A pair of quantities ($E_S, \sigma_S$) was extracted from the energy spectra – the mean value of the energy of the EM shower and the variance of its distribution. The energy resolution of the module was determined as the ratio of the standard deviation to the mean value of energy distribution. The $S$ index refers to the active part of the calorimeter, to the scintillator.

Parameters of the studied modules in units of radiation length $H[X_0]$, total module length $H_x$, cm, the sampling fraction $F_{samp} = (E_S + E_{Pb})$ presented in the Tables 1 - 4.

*Table 1*

*The sampling fraction $F_{samp}$ for different $D_{Pb}$ thicknesses*

| $D_{Pb}$, mm | $X_0 \cdot \rho$, g/cm² | $X_0$, cm | $F_{samp}$ |
|---|---|---|---|
| 0.3 | 8.75 | 3.15 | 0.39 |
| 0.4 | 8.18 | 2.53 | 0.32 |
| 0.5 | 7.84 | 2.16 | 0.27 |

Due to energy leakages from the calorimeter module, the deposit energy distribution differs from the Gaussian distribution. Our major challenge in the mathematical processing of simulated data is the correct extraction parameters from the fitting procedure.

The choice of function with the best approximation converge has received considerable attention. Previously, for asymmetric total absorption detector peaks, the Crystal Ball function (CBF) was used [6, 7]. But as shown in [8], this function has poor convergence and the obtained parameter values are highly dependent on statistics. Therefore, we used Das approximation [8], devoid of these shortcomings. This made it possible to use with great efficiency almost all the modeled input energy points in the maximal possible fit range, where the $\chi^2$ per degree of freedom is about one.

The procedure for extracting data from simulated EM showers was carried out using ROOT6 Data Analysis Framework [9] with options Minimizer ("Minuit2", "Migrad") in the entire energy range.

### 1.2. INCLUSION OF THE ENERGY LEAKAGE FROM CALORIMETER TO THE HIS RESPONSE

It is generally assumed that the shower is totally contained in a single tower. Coefficients *a* and *b* are found as a result of the approximation procedure sets of pairs of quantities ($E_S$, $\sigma_S$) according to the (1). In turn, the pair of quantities is obtained as a result of approximation the energy distribution of the calorimeter response to the input beam energy $E_0$. In general, the situation is more complicated. If part of the energy leaves in the form of leakages, then the energy balance $E_0 \neq E_S + E_{Pb}$ is disturbed.

Therefore, there is no linear correspondence between the input beam energy $E_0$ and absorbed energy in the module (Fig. 1). The longitudinal parameters of the module are shown in Fig. 1 as the number of pairs of plates $L$, and in centimeters and radiation lengths (see Tables 2 - 4).

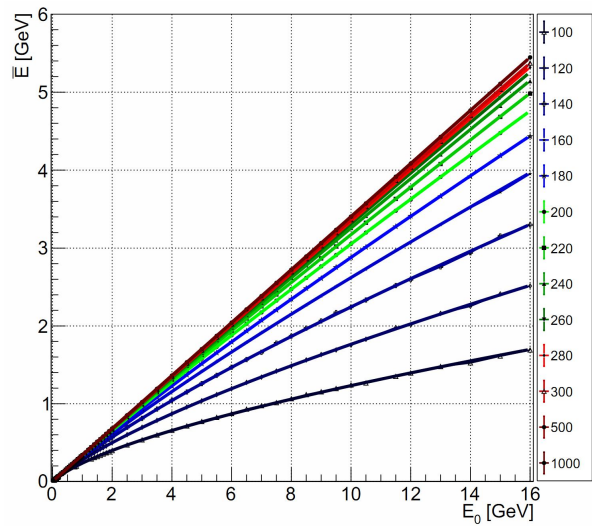

*Fig. 1. The central module response in 3×3 assembly on the electron beam energy. $H_{YZ}$=110 mm, $D_{Pb}$=0.3 mm*

It is clearly seen that the nonlinearity of the response is significantly manifested when significant energy leakages from the module. But even with small leakages, this will affect the shape of the energy resolution. The width of the absorption peak increases and the average energy of the electromagnetic shower decreases.

At large leakages, the asymmetry of the total shower absorption peak is already noticeable in the form of the left tail of the distribution (Fig. 2). Energy leakages are also energy dependent. For any real length of the module, leakages can be quite large at high energies. Longitudinal leakages from the module can also be from the front end of the module, since the beam always enters the module from the outside. They are insignificant, but there is always as albedo.

In 3×3 module assembly lateral leakages exist even for typical sizes $H_{YZ}$=110 mm² module KOPIO [1 - 3] with different amounts L. The spectra in the lead absorber, scintillator and EM leakages for modules of different transverse sizes (see Fig. 2). For instance, at $E_0$=1.0 GeV one can see, about 15% of the energy of the EM shower is absorbed by the module with $H_{YZ}$=110 mm and about three times less with $H_{YZ}$=300 mm. Accordingly, the energy resolution will also be different.

This is due to the fact that the wide module collects almost all the energy of the shower. Therefore, leakages can be considered as a kind of sampling fluctuations that occur at the border of the calorimeter module and the external environment. Longitudinal leakages occur at the module-air border and lateral leakages at the module and the adjacent module border. The approximation function will differ from the function (1), and the constant term should already depend on energy. We note an important detail, which is usually omitted, that the coefficients *a* and *b* are obtained as a result of approximation of the entire data set. These points are a collection of all responses to the input energy in the entire studied energy range. Therefore, even small leakages will affect the shape of the energy resolution function.

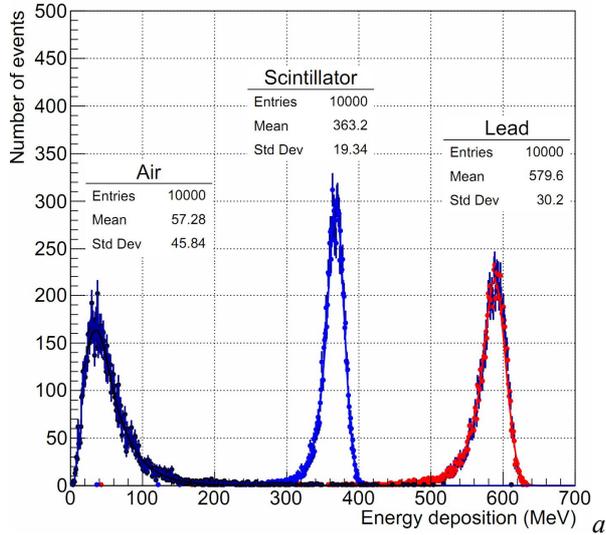

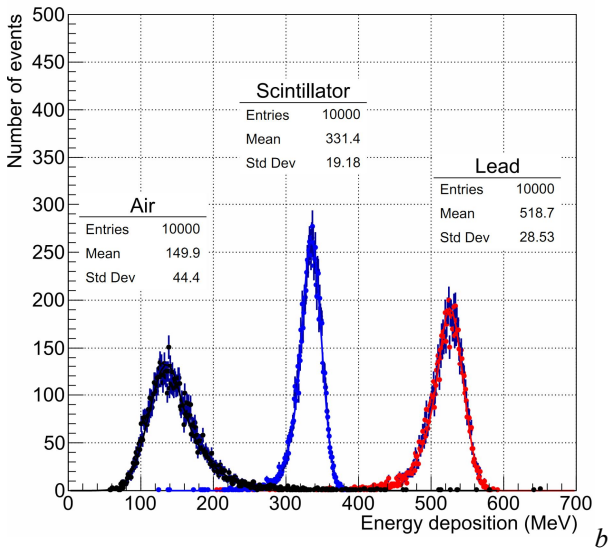

*Fig. 2. Calorimeter response at $E_0 = 1\ GeV$ for the module: L=220 ($H_X$=17$X_0$), $D_{Pb}$=0.3 mm. $H_{YZ}$=300 mm (a); $H_{YZ}$=110 mm (b)*

In a real experiment the electron energy reconstruction is determined in a fixed number of calorimeter cells are added together ("cluster"). Then the total absorption energy of the shower can be corrected.

## 2. METHOD AND ANALYSIS

When approximating the simulated data, we kept the form of the function (1), since the two terms in it have a generally accepted physical interpretation. Then still, the coefficient *a* determined by sampling of the calorimeter, and the coefficient *b* determined by EM leakages from the sandwich module structure. But the collection of the deposit energy in the module will be incomplete due to leakages and will naturally affect the coefficient *a*.

The authors of [10] tried to apply an approximation (2) for electromagnetic showers with leakages:

$$\frac{\sigma_E}{E} = \frac{a}{\sqrt{E}} \oplus b \oplus \left( p_1 \cdot f(E) + p_2 \cdot f^2(E) + p_3 \cdot f^3(E) \right), \quad (2)$$

where $f(E) \equiv \ln(E/E_c)$ – energy dependent function and $E_c$ – effective critical energy. The dependence of the constant term from the energy is represented by the first terms of the logarithmic series with parameters $p_1$ - $p_3$. These terms are determined from Grindhammer-Peter's parameterization [11] for description fluctuations due to energy leakage.

It is important to note here that in this approach the constant part *b* is explicitly allocated from the function. This part *b* will really not depend from energy even in the case of an ideal calorimeter. The other part determines the degree of influence of leakage on the energy resolution curve.

The data obtained from Monte Carlo simulation are presented as a family of energy resolution curves (Fig. 3, $D_{Pb}$=0.3 mm).

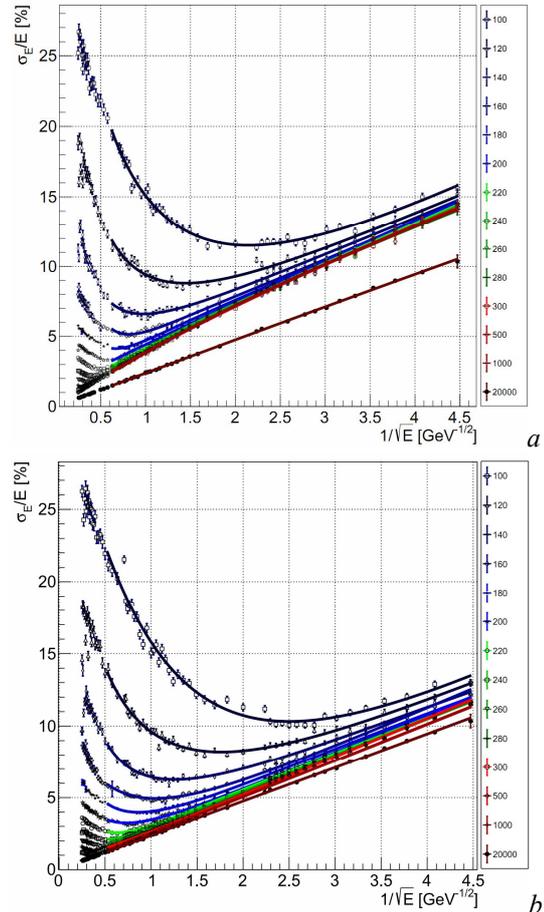

*Fig. 3. Energy resolution of calorimeters for the module $D_{Pb}$=0.3 mm. $H_{YZ}$=300 mm (a); $H_{YZ}$=110 mm (b)*

The upper Fig. 3 shows the curves for a transverse width module $H_{YZ}$=110 mm, and on the bottom – with transverse width $H_{YZ}$=300 mm. From the analysis of the figures, two areas of leakage are clearly visible. The first area is roughly 0…250 MeV (in Fig. 3: ~ 2…4.5 GeV$^{-1/2}$). This is the area of parallel rise of curves in the area of the main influence of the stochastic term relative to the reference line ($L = \infty$) for the unlimited size module.

In this way, because of the limited lateral size of the calorimeter module, an incomplete collection of EM shower occurs and the resolution deteriorates. Physically, this means that energy fluctuations in a wide modulus ($H_{YZ}$=300 mm) less, than in narrow module ($H_{YZ}$=110 mm). The second area is located approximately in range 250 MeV…16 GeV (in Fig. 3: ~ 0.24...2 GeV$^{-1/2}$). Here's the main influence have longitudinal leakage due to the release of shower energy from the module and albedo from the front end.

Approximation (2) has satisfactory convergence up to $\chi^2/NDF \sim 1$. The continuous curves in the figures extend up to about 4 GeV.

When fitting into the region of high energies, $\chi^2/NDF$ value deteriorates. It is required to involve additional terms of higher orders, which complicates their physical interpretation.

We also used a new approximation function which is similar to (1). This fitting function explicitly takes into account energy leakage over the entire energy range:

$$\frac{\sigma_E}{E} = \frac{a}{E^c} + b \cdot E^d, \quad (3)$$

where the coefficients $c, d \geq 0$. In contrast with function (2), this function has good convergence for all versions of the ideal calorimeter prototypes we have considered. Minor differences from $\chi^2/NDF \sim 1$, in our opinion, associated with an increase in systematic errors by algorithms ROOT6 Minimizer, which are difficult to control for all energy spectra. Formula (3) coincides with formula (1) in the limiting case ($c = 1/2$, $d = 0$), when leakages can be neglected.

Tables 2 - 4 show the final results for option $D_{Pb}$=0.3 mm. For comparison, the EM leakage $Lk$, % at $E_0 = 1$ GeV is also given.

Similar results were obtained for options $D_{Pb}$=0.4 mm and $D_{Pb}$=0.5 mm.

*Table 2*

*The final results for «cluster», only central module ($H_{yz}$=110)*

| L | 100 | 120 | 140 | 160 | 180 | 200 | 220 | 240 | 260 | 280 | 300 | 500 | 1000 | ∞ |
|---|---|---|---|---|---|---|---|---|---|---|---|---|---|---|
| $H_x$, cm | 18.0 | 21.6 | 25.2 | 28.8 | 32.4 | 36.0 | 39.6 | 43.2 | 46.8 | 50.4 | 54.0 | 90.0 | 180.0 | ∞ |
| $H_x$, $X_0$ | 5.7 | 6.9 | 8.0 | 9.1 | 10.3 | 11.4 | 12.6 | 13.7 | 14.8 | 16.0 | 17.1 | 28.5 | 57.1 | ∞ |
| a, %·GeV$^c$ | 0.05 | 1.02 | 2.29 | 3.19 | 3.59 | 4.04 | 4.14 | 4.11 | 4.01 | 4.04 | 4.01 | 3.97 | 3.96 | 2.39 |
| c | 1.67 | 0.82 | 0.63 | 0.53 | 0.50 | 0.46 | 0.45 | 0.46 | 0.47 | 0.46 | 0.47 | 0.47 | 0.47 | 0.50 |
| b, %·GeV$^{-d}$ | 18.65 | 11.88 | 6.89 | 3.77 | 2.13 | 0.99 | 0.47 | 0.26 | 0.19 | 0.08 | 0.04 | 0.0 | 0.0 | 0.0 |
| d | 0.16 | 0.23 | 0.31 | 0.38 | 0.43 | 0.53 | 0.63 | 0.64 | 0.59 | 0.73 | 0.79 | - | - | - |
| Lk, %, 1 GeV | 42.9 | 32.5 | 25.2 | 20.3 | 17.5 | 15.6 | 14.4 | 13.7 | 13.3 | 13.1 | 12.9 | 12.5 | 12.5 | 0.0 |

*Table 3*

*The final results for «cluster», only central module ($H_{yz}$=300)*

| L | 100 | 120 | 140 | 160 | 180 | 200 | 220 | 240 | 260 | 280 | 300 | 500 | 1000 | ∞ |
|---|---|---|---|---|---|---|---|---|---|---|---|---|---|---|
| $H_x$, cm | 18.0 | 21.6 | 25.2 | 28.8 | 32.4 | 36.0 | 39.6 | 43.2 | 46.8 | 50.4 | 54.0 | 90.0 | 180.0 | ∞ |
| $H_x$, $X_0$ | 5.7 | 6.9 | 8.0 | 9.1 | 10.3 | 11.4 | 12.6 | 13.7 | 14.8 | 16.0 | 17.1 | 28.5 | 57.1 | ∞ |
| a, %·GeV$^c$ | 0.0 | 0.32 | 1.33 | 2.11 | 2.44 | 2.60 | 2.60 | 2.65 | 2.57 | 2.47 | 2.40 | 2.38 | 2.38 | 2.39 |
| c | 2.53 | 1.06 | 0.68 | 0.56 | 0.52 | 0.51 | 0.51 | 0.50 | 0.51 | 0.52 | 0.54 | 0.54 | 0.54 | 0.50 |
| b, %·GeV$^{-d}$ | 18.86 | 12.77 | 7.79 | 4.53 | 2.74 | 1.61 | 1.02 | 0.61 | 0.45 | 0.39 | 0.34 | 0.15 | 0.15 | 0.0 |
| d | 0.16 | 0.21 | 0.27 | 0.34 | 0.38 | 0.43 | 0.44 | 0.47 | 0.42 | 0.33 | 0.26 | 0.01 | 0.0 | - |
| Lk, %, 1 GeV | 40.1 | 28.5 | 19.7 | 13.7 | 9.5 | 6.6 | 4.8 | 3.4 | 2.6 | 2.0 | 1.7 | 0.8 | 0.8 | 0.0 |

*Table 4*

*The final results for «global cluster» ($H_{yz}$=600)*

| L | 100 | 120 | 140 | 160 | 180 | 200 | 220 | 240 | 260 | 280 | 300 | 500 | 1000 | ∞ |
|---|---|---|---|---|---|---|---|---|---|---|---|---|---|---|
| $H_x$, cm | 18.0 | 21.6 | 25.2 | 28.8 | 32.4 | 36.0 | 39.6 | 43.2 | 46.8 | 50.4 | 54.0 | 90.0 | 180.0 | ∞ |
| $H_x$, $X_0$ | 5.7 | 6.9 | 8.0 | 9.1 | 10.3 | 11.4 | 12.6 | 13.7 | 14.8 | 16.0 | 17.1 | 28.5 | 57.1 | ∞ |
| a, %·GeV$^c$ | 0.0 | 0.36 | 1.26 | 2.01 | 2.52 | 2.64 | 2.67 | 2.65 | 2.56 | 2.53 | 2.46 | 2.39 | 2.40 | 2.33 |
| c | 2.39 | 1.00 | 0.68 | 0.55 | 0.49 | 0.48 | 0.47 | 0.47 | 0.48 | 0.49 | 0.50 | 0.50 | 0.50 | 0.50 |
| b, %·GeV$^{-d}$ | 19.00 | 12.76 | 7.84 | 4.50 | 2.54 | 1.50 | 0.85 | 0.52 | 0.37 | 0.23 | 0.19 | 0.02 | 0.01 | 0.06 |
| d | 0.17 | 0.21 | 0.28 | 0.35 | 0.42 | 0.46 | 0.51 | 0.51 | 0.47 | 0.49 | 0.41 | 0.0 | 0.0 | 0.0 |
| Lk, %, 1 GeV | 40.2 | 28.4 | 19.8 | 13.6 | 9.35 | 6.4 | 4.5 | 3.1 | 2.3 | 1.6 | 1.2 | 0.4 | 0.4 | 0.0 |

## CONCLUSIONS

In this paper, it is shown that the correct value of the stochastic and constant terms in the fitting formula for the energy resolution of an ideal sampling calorimeter of a module of real dimensions can be obtained only when energy leakage from the module are taken into account. Explicit consideration of energy leakage makes it possible to determine the constant term more adequately, taking into account the selection of points according to the criterion $\chi^2/ndf \sim 1$.

The basis of the applied approach is a rigorous selection of the EM showers obtained by the Monte Carlo simulation in the calorimeter in a narrow range of values $\chi^2/NDF$. To approximate the spectra, we used the function [8], which is much better at describing the asymmetrical shape of the spectrum with allowance for leakage than the CBF function. The energy resolution function is obtained from the spectra of the average energy values calculated in this way, as well as the values of the root-mean-square deviations. It is shown that the constant term does not depend on energy only in the absence of leakages. In this case, it is zero. This can be seen from the approximation (3) and from the simulation results. A family of energy resolution curves for possible prototypes of the ideal calorimeter of the SPD setup for the thicknesses of the lead absorber $D_{Pb}$ = 0.3, 0.4, 0.5 mm was obtained.

**ВЛИЯНИЕ УТЕЧЕК ЭНЕРГИИ НА ЭНЕРГЕТИЧЕСКОЕ РАЗРЕШЕНИЕ ЭЛЕКТРОМАГНИТНЫХ СЭМПЛИНГ-КАЛОРИМЕТРОВ**

*О.П. Гаврищук, В.Е. Ковтун, Т.В. Малыхина*

Рассматриваются прототипы идеального электромагнитного сэмплинг-калориметра установки SPD коллайдера NICA. С помощью моделирования детально исследуется влияние утечек энергии из модуля на его энергетическое разрешение. Получены величины стохастического и константного членов для различных вариантов прототипа калориметра SPD.

**ВПЛИВ ВИТОКІВ ЕНЕРГІЇ НА ЕНЕРГЕТИЧНУ РОЗДІЛЬНУ ЗДАТНІСТЬ ЕЛЕКТРОМАГНІТНИХ СЕМПЛІНГ-КАЛОРИМЕТРІВ**

*О.П. Гаврищук, В.Є. Ковтун, Т.В. Малихіна*

Розглядаються прототипи ідеального електромагнітного семплінг-калориметра установки SPD колайдера NICA. За допомогою моделювання детально досліджується вплив витоків енергії з модуля на його енергетичну роздільну здатність. Отримано величини стохастичного і константного членів для різних варіантів прототипу калориметра SPD.